\documentclass[aps,prd,10pt,a4paper,altaffilletter,amssymb,showpacs,nofootinbib,twocolumn,showkeys,a4paper,floatfix]{revtex4-1}
\usepackage{graphicx,epsfig}
\usepackage{subfig}
\usepackage{amsmath}
\usepackage{color}
\usepackage[normalem]{ulem}
\usepackage{natbib}
\usepackage{multirow}

\usepackage{tikz}
\usetikzlibrary{fit}
\tikzset{highlight/.style={rectangle,rounded corners,fill=red!15,draw,fill opacity=0,thick,inner sep=0pt}}
\newcommand{\tikzmark}[2]{\tikz[overlay,remember picture,baseline=(#1.base)] \node (#1) {#2};}

\newcommand{\Highlight}[1][submatrix]{%
    \tikz[overlay,remember picture]{
    \node[highlight,fit=(left.north west) (right.south east)] (#1) {};}
}

\newcounter{qcounter}

\begin{document}

\title{Nonparametric reconstruction of the cosmic expansion with local regression smoothing and simulation extrapolation}
\author{Ariadna Montiel$^{1}$, Ruth Lazkoz$^{2}$, Irene Sendra$^{2}$, Celia Escamilla-Rivera$^{2}$ and Vincenzo Salzano$^{2}$}

\affiliation{$^{1}$ Departamento de F\'{\i}sica, Centro de Investigaci\'on y de Estudios Avanzados del I. P. N., Apartado Postal 14-740, 07000 M\'exico D.F., Mexico.\\
$^{2}$ Departamento de F\'{\i}sica Te\'orica e Historia de la Ciencia, Universidad del Pa\'{\i}s Vasco (UPV/EHU), Apdo. 644, E-48080, Bilbao, Spain.}

\begin{abstract}

In this work we present a nonparametric approach, which works on minimal assumptions, to reconstruct the cosmic expansion of the Universe.
We propose to combine a locally weighted scatterplot smoothing method and a simulation-extrapolation method.
The first one (\textit{Loess}) is a nonparametric approach that allows to obtain smoothed curves with no prior knowledge of the functional relationship between variables nor of the cosmological quantities. The second one (\textit{Simex}) takes into account the effect of measurement
errors on a variable via a simulation process.
For the reconstructions we use as raw data the Union2.1 Type Ia Supernovae compilation, as well as recent Hubble parameter measurements.
This work aims to illustrate the approach, which turns out to be a self-sufficient technique in the sense we do not have to choose anything by hand.
We examine the details of the method, among them the amount of observational data needed to perform the locally weighted fit which will define the robustness
of our reconstruction. In view of our results, we believe that our proposal offers a promising alternative for reconstructing global trends of cosmological data when there is little 
intuition on the relationship between the variables and we also think it even presents good 
prospects to generate reliable mock data points where the original sample is poor.

\end{abstract}


\maketitle

\section{Introduction}

The cosmic acceleration of the Universe has been confirmed by several independent observations including Type Ia Supernovae (SNe Ia), the cosmic microwave background (CMB)
and the large scale structure (LSS) of the Universe \citep{Riess:1998cb,Perlmutter:1998np,Knop:2003iy,Riess:2004nr,Astier:2005qq,Spergel:2003cb,Spergel:2006hy,Tegmark:2003ud}. 
Typically, this accelerated expansion has been attributed to the existence of a new entity called dark energy (DE) which makes up nearly $68.6\%$ of the cosmic substratum but still
with unknown properties \citep{Planck}. Therefore, elucidating what drives the accelerated expansion of the Universe or establising the properties of dark energy are real challenges 
in cosmology.

The community has proposed a huge amount of theoretical scenarios that attempt to explain this recent acceleration of the Universe: $\Lambda$CDM
\citep{Friedmann1924, Lemaitre1927, Sitter1917}, quintessence \citep{Zlatev1999, Sahni2004},  Chaplygin gas
\citep{Kamenshchik:2001cp,Bento2002}, modified gravity \citep{Nojiri2006}, holographic dark energy \citep{Li2004},
braneworld models \citep{Maartens2004}, $f(R)$ theories \citep{Sotiriou2010}, theories with extra dimensions \citep{Lechtenfeld2002},
and quite a few others. However, despite their great compliance with observational data, none of them has provided  a conclusive answer about the nature of the DE.

This situation has motivated the study of other methods that can make the most of the observational data and give as much information as possible about
the properties of the dark energy. In general, these approaches attempt to reconstruct the properties of the DE or the history of the expansion rate as directly as possible
from observations, not establishing an association with a fundamental physical model. 
They  can be broadly classified into \textit{parametric} and \textit{nonparametric} methods. Parametric methods are viable approaches when the relationship between the variables of 
the phenomena
under study is known, and their goal is to constrain the parameters of the chosen model. Refs.~\citep{Chevallier2001,Linder2003,Wang2008a,Davis07,Serra07,Linder08,Soll09,Kil10,Alam2004} 
can be checked for details of data analysis and methods of parametric reconstruction of the properties of dark energy.  However, when there is no clue about the explicit form of the 
relationship between the variables or the functional form for the quantity of interest, one has to 
propose it, which can lead to misleading results. At this point is where nonparametric methods make their way into the scene. They try to provide the general trend of the variable of 
interest when the relationship between the variables is unknown or there is little intuition about it because the data do not have a clear interpretation. Indeed, they have become 
popular given their usefulness for enhancing scatter plots and other diagnostic plots with the goal of displaying the underlying structure in the data \citep{Cook99, Fox08}.

In the literature one can find several approaches covering nonparametric and model independent reconstructions \citep{Huterer2003,Clarkson2010,Tang08,Critten05,Cri11,Daly03,Daly04,Daly08,Lazkoz12,Espana05,Fay2006,
Bonvin06,Shafieloo06,Shafieloo07,Shafieloo10,Shafieloo12,Bogdanos09,Nesseris10,Holsclaw:2010nb,
Holsclaw:2011wi,Holsclaw2010,Seikel2012,Shafiel12,Seikel12,Yahya,BenitezH11,Benitez13,AlbertoVazquez2012}, although, most of them must deal with the scarceness of data or some other limitation intrinsic to the method. Such approaches include the Principal Components Analysis (PCA) \citep{Huterer2003}, the Nonlinear Inverse Approach (NIA) \citep{Espana05}, the Dipole of the Luminosity Distance method (DLD) \citep{Bonvin06}, the Smoothing Method (SM) \citep{Shafieloo06}, Gaussian Processes (GP) \citep{Holsclaw2010}, Nodal Reconstruction (NR) \citep{AlbertoVazquez2012}, Genetic Algorithms (GA) \citep{Bogdanos09} and three representative approches of Model Independent Reconstructions of the Expansion History (MIR-I), (MIR-II) and (MIR-III), corresponding to the schemes presented in \citep{Daly03}, \citep{Fay2006} and \citep{BenitezH11}, respectively.

Even though each one of the above methods are well established, none of them provides a totally compelling procedure within which the accelerated expansion or the nature of the DE can be understood. In this context, we can point out some features and shortcomings that they present in common:

\begin{list}{\roman{qcounter}.}{\usecounter{qcounter}}
\item The assumption of a prior, a fiducial cosmological model or a initial guess model, which leads to biased results, as happens in GP, SM, NIA, PCA, NR, MIR-II and MIR-III.

\item A binned approach in which the bins share data points. It causes fluctuations when the individual data points enter and leave a fitting window, as happens in MIR-I.

\item Low efficiency at high redshifts or in regions with few data points, as happens in MIR-I, NIA, DLD, PCA, and NR. In most of cases, this can be solved adding more data points, 
however this can result in computational issues or in numerical instability.

\item Underestimation of the error or the absence of tools to estimate and propagate errors are suffered from GP, SM, GA and PCA.

\item GP, SM, NIA, PCA seem to suffer from a high computational cost given by the method itself or by the number of data.

\end{list}

Convinced as we are that this line of investigation can give some guidance to elucidate the nature of the DE or of the entity that drives the accelerated expansion, and additionally, 
motivated by the fact that the approaches proposed have not succeded in their attempt to reconstruct the cosmic expansion, we propose to perform such reconstruction
using a locally weighted scatter plot smoothing method (\textit{Loess}),  which overcomes these difficulties. 
Since \textit{Loess} is a nonparametric method, 
we do not have to assume any functional form of the statistical relationship between the variables, 
and the functional form is estimated from the raw data. Besides, it is a completely cosmological-model-independent 
method because it does not require the input of any cosmological model nor any information concerning cosmological parameters. 
Although \textit{Loess} relies on bins, it does not have a similar problem to the one suffered by MIR-I, because it is a locally weighted fit. 
However, \textit{Loess} on its own does not take into account the 
measurement error of the observations to perform the reconstruction, that is why we propose to combine it with 
a simulation-extrapolation method (\textit{Simex}), which addresses the effects of the measurement errors on parameter estimates. 
Thus, \textit{Loess} + \textit{Simex} turns out to be a very simple approach that provides successfully the global trend of the data with a very low computational cost, 
besides it is applicable with the same 
efficiency in the whole redshift range and can estimate and propagate the error
thanks to the analogy with some properties of parametric approaches. Moreover,  the reconstruction can be used to infer
mock data points (through the local polynomial) where the original sample is poor, and yet again this is done cosmological-model-independently, just inheriting the global trend.

In this paper we present the most important features of the method, which appears as a promising alternative to reconstruct the cosmic expansion. 
The work is structured as follows: in Sec. \ref{basics}, we introduce the basics of \textit{Loess}; in Sec. \ref{simex} the logic behind \textit{Simex}; next, in Sec. \ref{cosmicexp}, we present the steps that need to be followed to apply their combination to astronomical observational data to obtain the cosmic expansion; in Sec. \ref{dataresults}, we detail the observational data samples chosen for our analysis; in Sec. \ref{discussion} we present and discuss the principal results and finally, in Sec. \ref{conclusions}, we discuss our concluding remarks.

\section{Basics of Simulation Extrapolation and Nonparametric Regression Method }

\subsection{Locally Weighted Scatterplot Smoothing (\textit{Loess})\label{basics}}

Locally weighted scatterplot smoothing (also known as local polynomial regression), originally introduced in \citep{Cleveland79} and further developed in \citep{Cleveland88}, is a generalization of standard least-squares methods for data analysis. It has become the most commonly used method for nonparametric simple regression in some disciplines. \textit{Loess} is a 
nonparametric method in the sense that the fitting is performed without having to specify in advance the relationship between the dependent and independent variables.

The procedure tries to depict the global trend of a dataset formed by $n$ observational measurements of a certain response $y_{i} \equiv y(x_{i})$, where $i =1, \ldots, n$, 
corresponding to certain values of the predictor or independent variable. One has to focus initially on the $i$th measurement, given by the pair $(x_{i},y_{i})$. A low degree polynomial is chosen as the regression 
function that will give us an approximation to the response, called $\hat{y}_{i} \equiv \hat{y}(x_{i,0})$, in a range of predictor values or points around the focus points 
$x_{i} \equiv x_{i,0}$\footnote{The suffix 0 is used here to stress the role that each $x_{i}$ data point has in the fitting and windowing procedure. In particular, $x_{i,0}$ 
will symbolize the $x_{i}$ point, chosen in turn to be the \textit{center} of the fitting window.}. The process is repeated so the whole range of $i$ is covered. Note that in this work, when we refer to $y$, $\hat{y}$ and $x$, we indeed refer to 
original $H(z)$ or $\mu(z)$ data ,  
simulated $H(z)$ or $\mu(z)$ data and the redshift $z$, respectively.

The subset of data centered at $x_{i,0}$  of length $m<n$ (see below), will be chosen by the nearest neighbors rule with the help of weights according to a Kernel (see further below). The fit is performed
using weighted least squares; specifically, more weight is given to points near the point whose response is being estimated, and less weight to points further away. Typically, 
local polynomials to fit each subset of data are of first or second order. Higher orders are possible, but do no really improve the final result and rather slow down the process
computationally. Eventually, this whole process offers the possibility to get the full view of the global trend of the data, which was the original objective of the procedure. 
To do so we simply have to join the reconstructed points with a line, thus obtaining a graphical account of the relationship between dependent and independent variables.

In the following we will explain briefly the important features of the method: the selection of number of data used in each fit, the degree of the polynomial and the form of the 
weight function. Additionally, we will address how to construct confidence intervals around a \textit{Loess} curve.

\subsubsection{Smoothing Parameter $\&$ Window Width}

The first step is to determine how many data points should be used in each weighted least squares fit. This is done through the \textit{smoothing parameter}, $s$, also called \textit{span},
which ranges between $0$ and $1$ and controls the flexibility of the \textit{Loess} regression function. When large values of $s$ are chosen, a large number of data points are used to fit 
and smoothest functions are produced
with a lower response to fluctuations in the data. 
On the other hand, using small values of the smoothing parameter $s$ means to fit a low number of data points, thus producing more irregular reconstructed curves, because the intrinsic 
noise and dispersion of data is fully captured.

The election of the span can be done roughly by trial and visual inspection of the effects of different values of $s$ on the global trend. Here instead, the election of the optimal value 
of the span $s$ will be done by using \textit{cross-validation} \citep{Fox08}, a more formal method to estimate and to select the best smoothing parameter $s$. The basic idea behind this algorithm is to estimate the mean-squared error of the fit \citep{Fox08,GVK019307713}. The hope is then that the smoothing parameter minimizing this estimate is also a good estimate for the mean-squared error itself \citep{GVK019307713}. Basically, cross-validation consists in omitting the \textit{i}th observation from the local regression at the focal value $x_{i,0}$; the resulting estimate will be denoted by $\hat{y}_{-i}$. %$\hat{y}_{-i} \vert x_i$.
The cross-validation function is
\begin{equation}
CV(s)=\frac{1}{n}\sum_{i=1}^{n} \left(\hat{y}_{-i}(s) - y_i\right)^2,
\label{Eq:CV}
\end{equation}
where $\hat{y}_{-i}(s)$ is $\hat{y}_{-i}$ for span $s$. Note that omitting the \textit{i}th observation, the fitted value $\hat{y}_{-i}$ is independent of the observed value $y_i$.

In practice, it is necessary to compute $CV(s)$ for a range of values of $s$. The value of $s$ that minimizes this function is considered to be the optimal amount of smoothing to apply to the local regression fit. Once the value of $s$ has been determined, $m = n \cdot s$ (rounded to the next largest integer) will give the number of data points that will have to be used in each weighted least squares fit.

\subsubsection{Weight function}

One key element of \textit{Loess} is the \textit{Kernel estimation}, such that in each fitting window containing $m$ data points, the fit gives more weight to
observations that are closer to the focal point $x_{i,0}$. The use of weights is supported by the guess that points near each other are, probably, more correlated to each other than points that are further apart. So that, following this logic, nearer points are likely to follow the same local model and may exert more influence on the estimation of the local parameters $\hat{y}_{i}$, while farther points are less likely to share the local model and may have less influence on the same estimates.

The weight or kernel function depends on the variable $\bar{x}\equiv (x_j-x_{i,0})/h$, the scaled distance between the predictor values for the \textit{j}th observation falling in the window with $x_{i,0}$ as the chosen focal point, with $j=1, \ldots, m$, and $h$ being the maximum distance between the point of interest and the $j-$elements of its window. By construction, after scaling the distance, the maximum absolute distance between the point of estimation and the farthest point in the window is $\lesssim 1$.

In this work, following standard practice, the weight function will be the tricube kernel:
\begin{equation}
K(\bar{x}) = \begin{cases}
  \left(1-\vert \bar{x} \vert^3 \right)^3 & \text{for $\vert \bar{x} \vert < 1$} \\
  0 & \text{for $\vert \bar{x} \vert \geq 1$}
\end{cases}\; ,
\label{eq:funw}
\end{equation}
and the weights used for the regression are $w_{ij}=K[(x_{j}-x_{i,0})/h]$. Of course, in each fitting window one has $0 < w_{ij} <1$ for the $m$ neighbours of $x_{i,0}$, 
and $w_{ij} = 0$ for all the other $n-m$ points.

Once the weights $w_{ij}$ have been calculated, we proceed to compute the fitted value at $x_{i,0}$ for the observed quantity $y_{i}$, i.e. we obtain the set of values $\hat{y}_{i}$.

The local polynomials that fit each subset of data are usually of first or second degree. Higher-degree polynomials are possible, and would work in theory, 
but would result in models that are not really compliant with the spirit of \textit{Loess}, which looks for a low-order polynomial and a simple model 
that can fit data easily. In this work, we shall consider that a linear polynomial is the most appropriate one to fit each subset of data.

%%%%%%%%%%%%%%%%%%%%%%%%
\subsubsection{Confidence bands around \textit{Loess} curve}

In a parametric regression, the central objects of estimation are the regression coefficients. However, in a nonparametric regression, like the one we are using here, there are no regression coefficients, 
and the central objective is the estimation of the regression function itself and its visualization, such that statistical inference focuses on the regression function directly.

To construct the confidence regions of the nonparametric regression we follow Ref. \citep{Fox20}. We start from the local polynomial estimate $\hat{y}_{i}$ that results from the locally weighted least-squares regression of $y$ on the $x$ values in each chosen window.

By assumption, the $y_i$'s are independently distributed, with common conditional variance $V(y_i)=\sigma^2$; %$V(y\mid x=x_i)=V(y_i)=\sigma^2$,
then, the sampling variance of the fitted value $\hat{y}_i$ is
\begin{equation}
\hat{V}(\hat{y}_i)=\sigma^2 \sum_{j=1}^{n} w_{ij}^2.
\end{equation}
However, to apply this result we require an estimate of $\sigma^2$. In linear least-squares simple regression the error variance is estimated by
\begin{equation}
S^2=\frac{1}{n-2}\sum_{i}^{n} r^2_i,
\end{equation}
where $r_i=y_i-\hat{y_i}$ corresponds to the residual for observation $i$, and $n-2$ to the degrees of freedom associated with the residual sum of squares. In a nonparametric regression, the residuals can be computed in the same way, $r_i=y_i-\hat{y_i}$, however, the degrees of freedom or number of parameters must be replaced by the \textit{effective degrees of freedom}, $df_{mod}$.

Once again we make the analogy with least squares regression. In this simple case, the way to determine the degrees of freedom is immediate, because the number of parameters is known, although a more precise and correct way would be to compute the trace 
of the hat matrix $\mathbf{H}$, which maps $\hat{y}$ into $y$ \citep{Andersen,Fox20,Cosma}. Despite the fact that in a nonparametric regression there are no parameters to sum, approximate degrees of freedom are obtained from the trace of the smoother matrix $\mathbf{S}$, which plays the same role as $\mathbf{H}$ in that it transforms $\hat{y}$ into $y$ \citep{Andersen,Fox20}. For kernel smoothers, which is indeed our case, $\mathbf{S}$ can be directly calculated from the kernel \citep{Cosma,Larry}, as the matrix of $w_{ij}$ elements.

There are other two popular definitions for the effective degrees of freedom: $df_{mod}=\mathrm{Tr} (\mathbf{S}\mathbf{S}^T) $, which we have adopted 
here by convenience without loss of generality; and $df_{mod}=\mathrm{Tr}(2\mathbf{S}-\mathbf{S}\mathbf{S}^T)$ \citep{Fox20}. For a least-squares fit, 
the two definitions involving $\mathbf{H}$ are equivalent; %the hat matrix $\mathbf{H}$ is a 
perpendicular projection operator, which is symmetric and idempotent and thus, $\mathrm{Tr} (\mathbf{H})=\mathrm{Tr} (\mathbf{H}\mathbf{H}^T)$.
however, for linear smoothers they can give two different results, even if they are often of similar magnitude \citep{Fox20,Handbook2}. 

It is worth mentioning that, unlike for linear parametric regression, the degrees of freedom for nonparametric one are not necessarily whole numbers \citep{Fox20}; and also that even though a nonparametric regression uses the equivalent of $df_{mod}$ parameters, this does not mean that if a global fit to data is performed using a $df_{mod}$-degree polynomial, it will produce the same regression curve \citep{Fox08}.

Once the $df_{mod}$ has been estimated, the residual degrees of freedom can be computed through $df_{res}=n-df_{mod}$, and the estimated error variance is finally given by:
\begin{equation}
S^2=\frac{1}{df_{res}}\sum_{i}^{n} r^2_i,
\end{equation}
whereas the estimated variance of the fitted value $\hat{y_i}$ is
\begin{equation}
\hat{V}(\hat{y}_i)=S^2\sum_{j=1}^{n} w_{ij}^2.
\end{equation}

Thus, assuming normally distributed errors, the $68\%$-percent confidence interval and the $95\%$-percent confidence interval of the regression function are approximately $\hat{y_i} \pm \sqrt{\hat{V}(\hat{y}_i)}$ and $\hat{y_i} \pm 2\sqrt{\hat{V}(\hat{y}_i)}$, respectively.

Although this procedure for constructing a confidence region has the virtue of simplicity, it is not completely rigorous, due to the bias in $\hat{y}_{i}$ as an estimate of the regression function. Such a bias in $\hat{y}_i$ can produce an over estimation of the error variance thus making the confidence interval too wide. %, or shift the same confidence interval at any point.
However, notice that as we have employed the cross-validation procedure to choose the optimal value of $s$, the bias that comes from the value of the span should be small.

Because $\hat{y}_{i}$ can be biased, strictly one should refer to the envelopes $\hat{y_i} \pm \sqrt{\hat{V}(\hat{y}_i)}$ and 
$\hat{y_i} \pm 2\sqrt{\hat{V}(\hat{y}_i)}$ around the sample regression as \textit{variability regions} rather than confidence regions. However, in this work we will 
regard them as confidence regions but having in mind this specification.

Up to this point, we have addressed briefly the features and free parameters of the \textit{Loess} method, however many 
more details of it can be found in Refs. \citep{Fox20, Andersen, Handbook,Cosma,Larry} and references therein.

\subsection{The simulation and extrapolation method (\textit{Simex})\label{simex}}

So far we have not used the observational errors $\sigma_{i}$ on real data $y_{i}$, because they are not contemplated by \textit{Loess} literature. 
To take them into account, we join the \textit{Loess} method with the \textit{Simex} one.

  \textit{Simex} is a simple simulation algorithm that allows to display the effect of measurement errors on parameter estimates. It was originally introduced in %Cook and Stefanski (1994-1995)
\citep{Cook94, Cook95} and has been applied in various fields but, to our knowledge, not in cosmology \citep{Carroll95,Ruppert04,Carroll99,Lin00}. We believe that it could be implemented with \textit{Loess} for fitting smooth curves to cosmological empirical data including measurement errors. In this way we could reconstruct the expansion history of the Universe with a high precision and without considering any prior on the cosmological quantities.

Here we provide a brief description, following Ref. \citep{ACM09}, of an adapted version of the \textit{Simex} algorithm, but further details of the method are available in Refs. \citep{Cook94, Cook95, Carroll06}. \textit{Simex} starts by taking each observation $y_i$ in the data set, with $i=1,...n$ and $n$ the number of data points, to which a known amount of measurement error is added as follows:
\begin{equation}
\eta_i(\lambda)=y_i+\sqrt{\lambda} \sigma_i, ~~~~ \lambda > 0
\label{Eq:simex}
\end{equation}
where $\sigma_i$ is the measurement error variance associated to the observed data $y_i$. A standard normal distribution of the errors is implicitly assumed. After introducing the variable $\lambda$, the final measurement error variance associated with the simulated data points, $\eta_i(\lambda)$, is $(1+\lambda) \sigma^{2}_{i}$ so that extrapolating its value to $\lambda \rightarrow -1$, 
we return back to the original data without uncertainties. We will refer to this scenario as error free situation.

The parameter $\lambda$ is actually a vector of length $\mathcal{N}$, a common choice  \citep{ACM09,Cook94} being $\lambda = \lbrace 0, 0.5, 1.0, 1.5, 2.0 \rbrace$. However, we have chosen to work, without loss of generality, with $\lambda = \lbrace 0.5, 0.6, 0.7,..., 2.0 \rbrace$ in order to have more simulated data sets. Then, Eq.~(\ref{Eq:simex}) is repeated for each data point $y_{i}$ and for each chosen $\lambda_{j}$ value, with $j=1, \ldots, \mathcal{N}$.

Thus finally we have that at each predictor $x_{i}$ a set of values $\eta_{i}(\lambda_{j})$ is attributed, which are obviously functions of the chosen $\lambda_{j}$ values. The $\lambda\rightarrow -1$ extrapolation required by \textit{Simex} theory will be then obtained after a standard regression of $\eta_{i}(\lambda_{j})$ is performed. A linear or quadratic polynomial are some possible choices \citep{ACM09}. In this work, we have found that the quadratic polynomial is the optimal choice
\begin{equation}
\eta(\lambda) =\beta_1+\beta_2 \lambda+\beta_3 \lambda^2 \, .
\label{Eq:polyn}
\end{equation}
The reconstructed $\eta_{i, \textit{Simex}}\equiv \hat{\eta}$, and the related confidence regions, can be obtained by taking $\eta(\lambda\rightarrow-1)$, thus 
taking back each data point to the error free situation.

\subsection{Joining \textit{Loess} and \textit{Simex} \label{cosmicexp}}% $H(z)$ and Supernovae Type Ia data}

\begin{figure*}
\begin{tabular}{cc}
\includegraphics[width=0.48\textwidth]{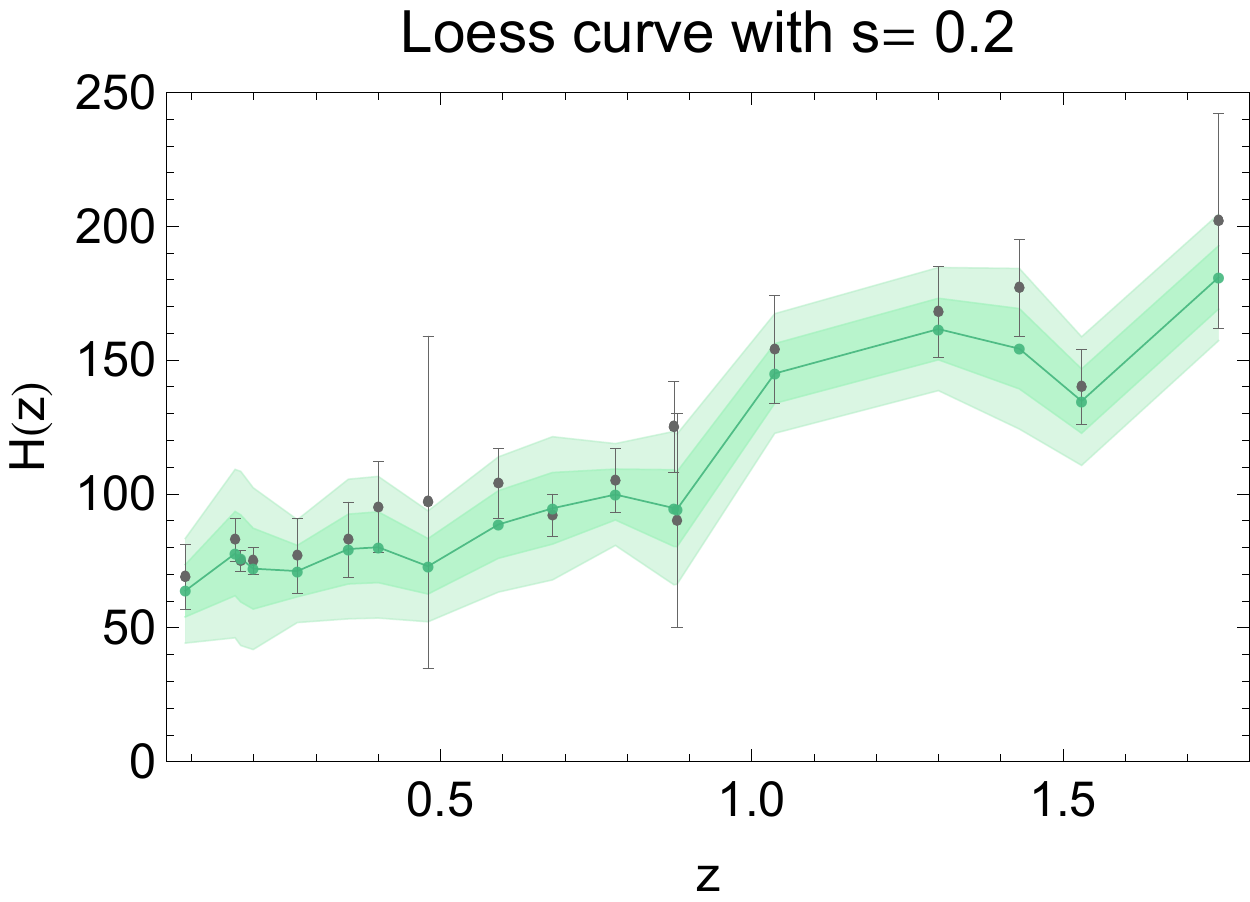}&
\includegraphics[width=0.48\textwidth]{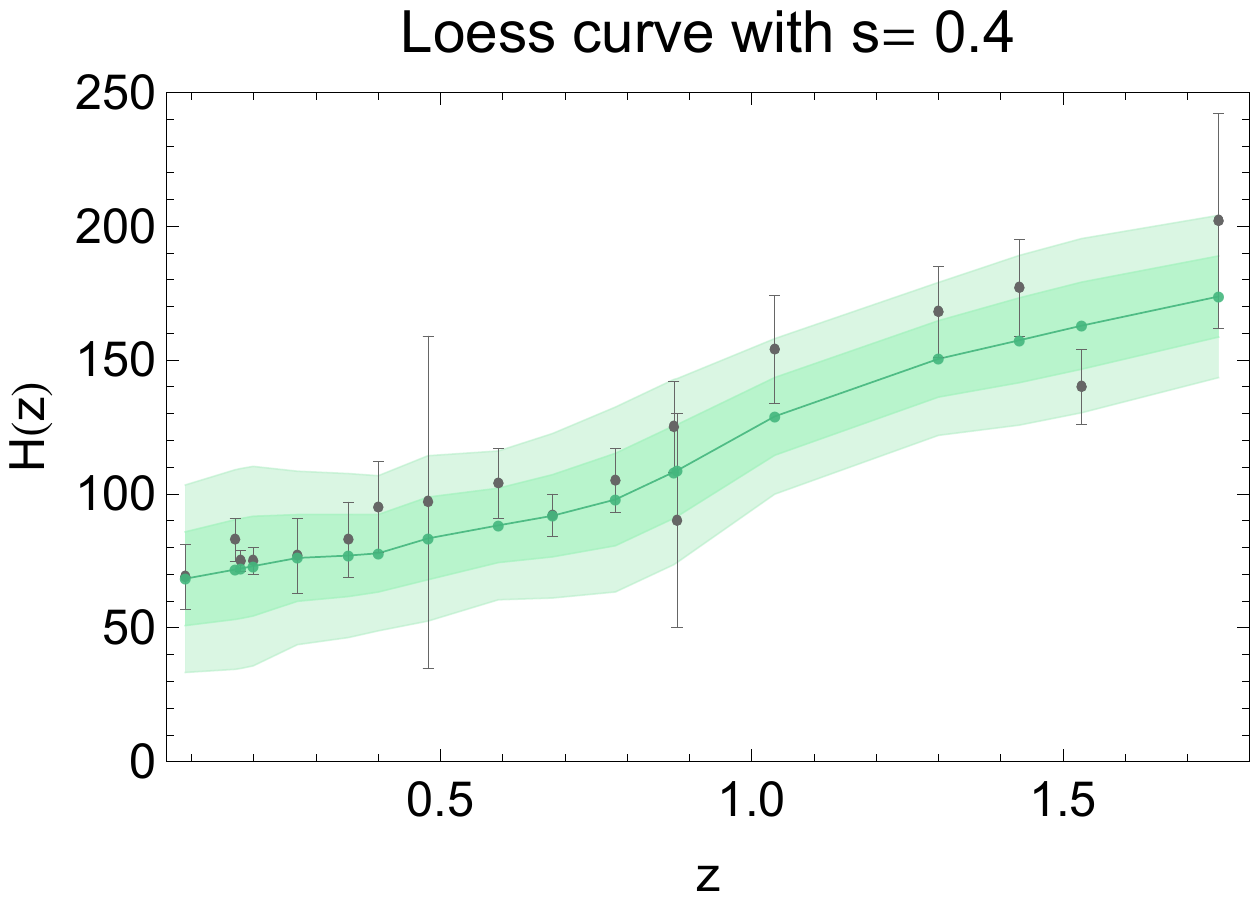}\\
%(a) & (b)
\end{tabular}
\begin{tabular}{cc}
\includegraphics[width=0.48\textwidth]{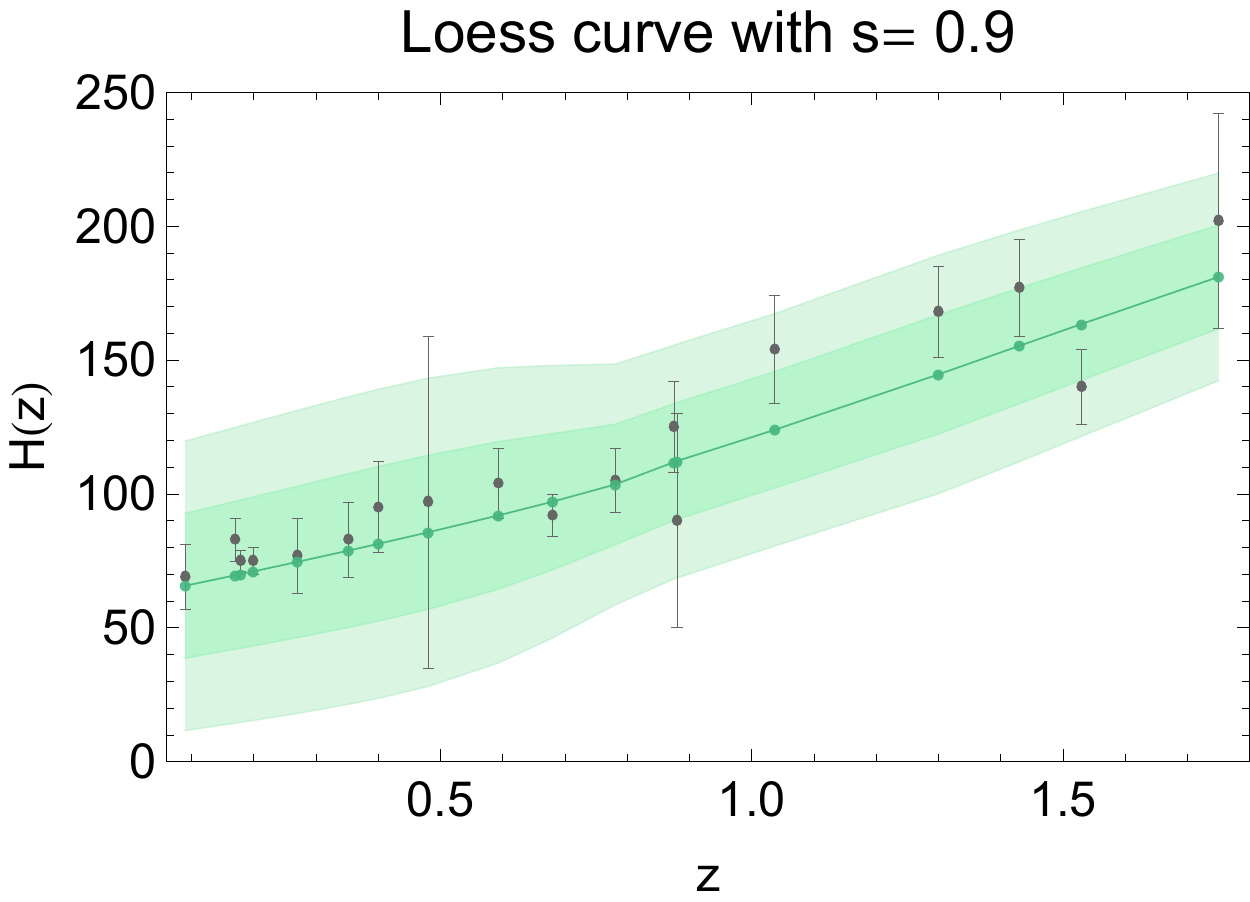}&
\includegraphics[width=0.48\textwidth]{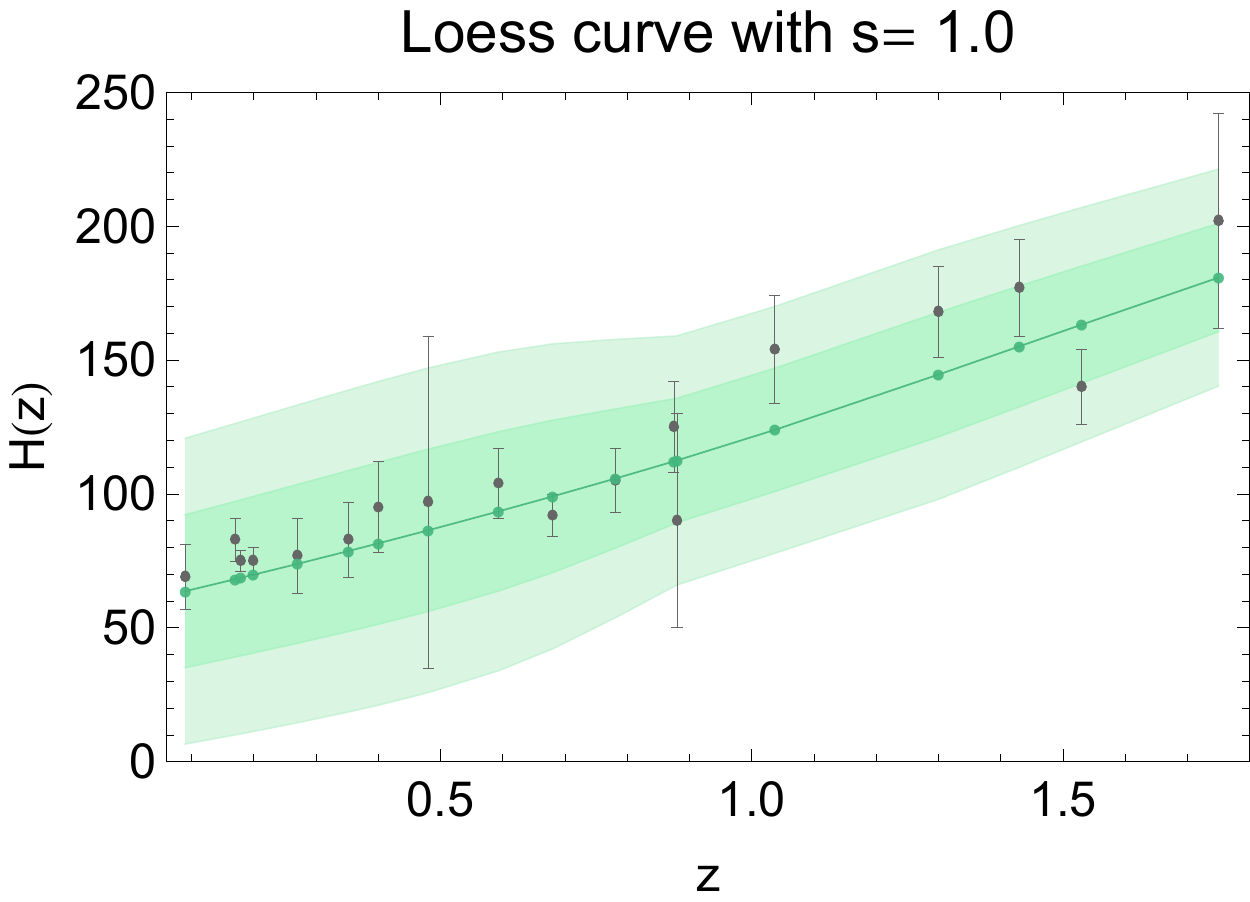}\\
%(c) & (d)
\end{tabular}
\caption{\label{Fig:Loess2} \textit{Loess} plots with different bandwidths. The gray points are the measurements of Hubble parameter including their uncertainties, 
the green points are the simulated data resulting from our \textit{Loess}+\textit{Simex} method. The central green line is obtained just connecting this dots and 
represents the best fit. The shaded contour represent $1\sigma -2 \sigma$ confidence level for our best fit.}
\end{figure*}

\begin{figure}
\includegraphics[width=0.4\textwidth]{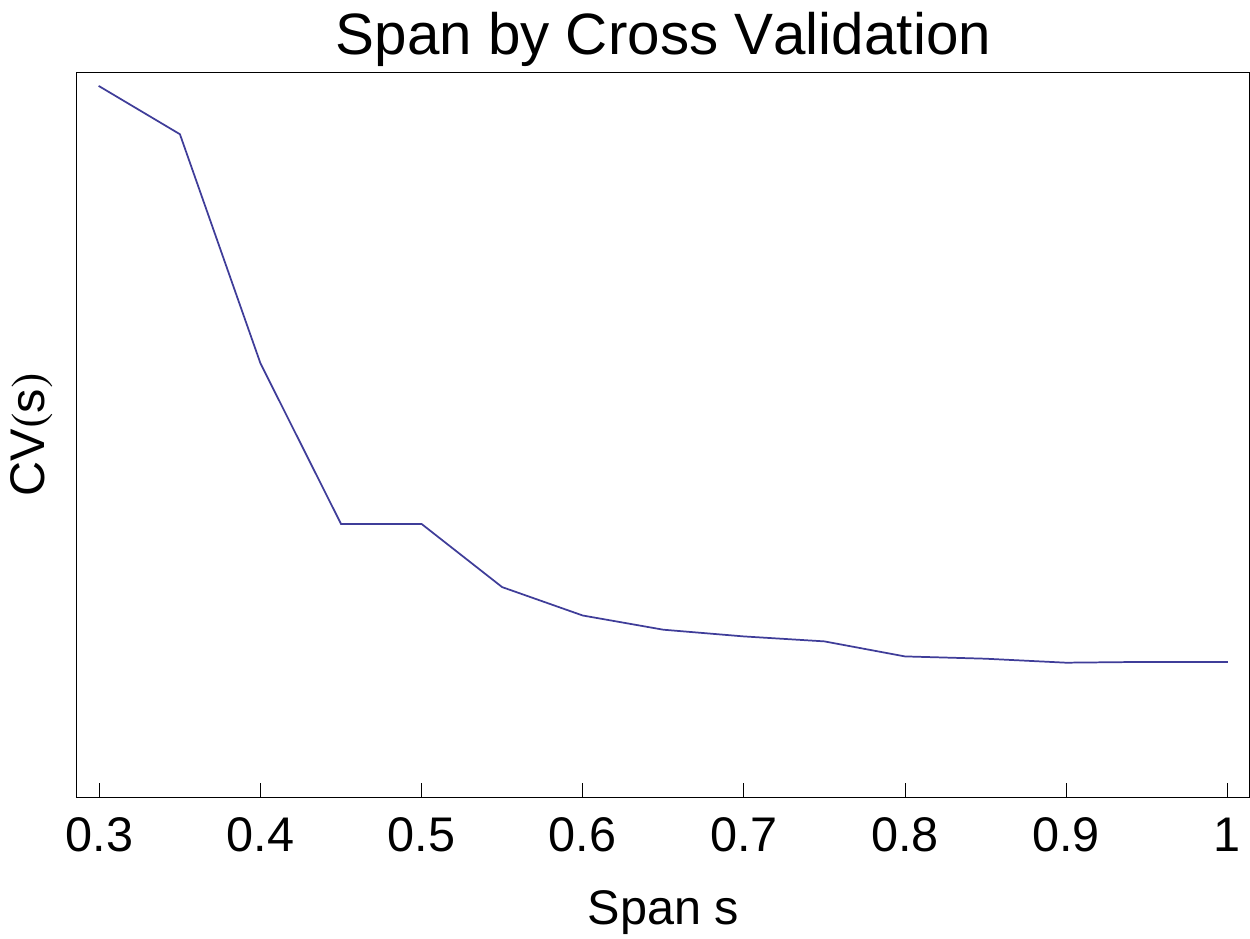}
\caption{\label{Fig:CVHz} Plot of $CV(s)$ versus $s$ for Hubble parameter. Good choices for $s$ goes from 0.8 until 1.0, although this 
last case could produce an over smoothed curve.}
\end{figure}

The main novelty of this work is to reconstruct the expansion history of the Universe using a combination of \textit{Loess} and \textit{Simex}. Both methods have been widely studied, and even utilized in various disciplines, although, to our knowledge, always been done independently. So what we look for through the implementation of both methods is in essence the global trend of the data, which in turn, could provide a clue of the most appropriate parameterization that must be between the variables.

Here we show schematically how we implement these methods:
\[
\left(
\begin{array}{*6{c}}
  y_{1} & y_{2} & \ldots\;\, & \tikzmark{left}{$y_{i}$}\tikzmark{right}{$y_{i}$} & \;\,\ldots & y_{n} \\
\end{array}
\right)
\Highlight[third]
\]
\tikz[overlay,remember picture] {
\draw[->] (4.25,0.4) -- (4.25,-0.4)
node [pos=0.5,left,xshift=-0.1cm] {$Simex$}
node [pos=0.5,right,xshift=0.1cm] { $(add \; \sqrt \lambda_{j} \sigma_{i})$};
}

%\vspace{0.05cm}

\[
\left(
\begin{array}{*6{c}}
  \eta_{1}(\lambda_1) & \eta_{2}(\lambda_1) & \ldots & \eta_{i}(\lambda_1) & \ldots & \eta_{n}(\lambda_1) \\
  \vdots              & \vdots              & \vdots & \vdots              & \vdots & \vdots              \\
  \tikzmark{left}{$\eta_{1}(\lambda_j)$} & \eta_{2}(\lambda_j) & \ldots & \eta_{i}(\lambda_j) & \ldots & \tikzmark{right}{$\eta_{n}(\lambda_j)$} \\
  \vdots              & \vdots              & \vdots & \vdots              & \vdots & \vdots              \\
  \eta_{1}(\lambda_{\mathcal{N}}) & \eta_{2}(\lambda_{\mathcal{N}}) & \ldots & \eta_{i}(\lambda_{\mathcal{N}}) & \ldots & \eta_{n}(\lambda_{\mathcal{N}})
\end{array}
\right)
\Highlight[first]
\]
\tikz[overlay,remember picture] {
\draw[->] (4.25,0.2) -- (4.25,-0.6)
node [pos=0.5,left,xshift=-0.1cm] {$Loess$}
node [pos=0.5,right,xshift=0.1cm] {($\eta \rightarrow \hat{\eta}$)};
}

\vspace{0.25cm}

\[
\left(
\begin{array}{*6{c}}
  \hat{\eta}_{1}(\lambda_1) & \hat{\eta}_{2}(\lambda_1) & \ldots & \tikzmark{left}{$\hat{\eta}_{i}(\lambda_1)$} & \ldots & \hat{\eta}_{n}(\lambda_1) \\
  \vdots              & \vdots              & \vdots & \vdots              & \vdots & \vdots              \\
  \hat{\eta}_{1}(\lambda_j) & \hat{\eta}_{2}(\lambda_j) & \ldots & \hat{\eta}_{i}(\lambda_j) & \ldots & \hat{\eta}_{n}(\lambda_j) \\
  \vdots              & \vdots              & \vdots & \vdots              & \vdots & \vdots              \\
  \hat{\eta}_{1}(\lambda_{\mathcal{N}}) & \hat{\eta}_{2}(\lambda_{\mathcal{N}}) & \ldots & \tikzmark{right}{$\hat{\eta}_{i}(\lambda_{\mathcal{N}})$} & \ldots & \hat{\eta}_{n}(\lambda_{\mathcal{N}})
\end{array}
\right)
\Highlight[second]
\]
\tikz[overlay,remember picture] {
\draw[->] (4.25,0.2) -- (4.25,-0.6)
node [pos=0.5,left,xshift=-0.1cm] {$Simex$}
node [pos=0.5,right,xshift=0.1cm] {($\lambda \rightarrow -1$)};
}

\vspace{0.25cm}

\[
\left(
\begin{array}{*6{c}}
  \hat{y}_{1} & \hat{y}_{2} & \ldots\;\, & \tikzmark{left}{$\hat{y}_{i}$}\tikzmark{right}{$\hat{y}_{i}$} & \;\,\ldots & \hat{y}_{n} \\
\end{array}
\right)
\Highlight[fourth]
\]

Explicitly the steps followed are:
\begin{enumerate}
\item Start \textit{Simex}:
 \begin{enumerate}
             \item Take as input the data points $y_{i}$. \label{sim1}
\item Select an $i$  value.
             \item Assign to each data point $y_{i}$ a certain $\lambda_{j}$ following Eq.~(\ref{Eq:simex}), thus obtaining the corresponding $\eta_i(\lambda_j)$, with $y_i$ standing for $H(z_i)$ or $\mu(z_i)$ and $\sigma_{y_i}$ for the corresponding measurement errors $\sigma_{H(z_i)}$ or $\sigma_{\mu(z_i)}$.
\item Go back to step 1(a) until all $i$ values are covered. {\it The vector of $y_i$ elements has become the matrix of
$\eta_i(\lambda_j)$ elements.}
\end{enumerate}
          \item Do \textit{Loess}:
                \begin{enumerate}
\item Select a $j$  value.
                      \item Assume the $\eta_{i}(\lambda_{j})$ values as our workable data instead of $y_{i}$. {This means we work with the elements of a row of
the matrix with $\eta_i(\lambda_j)$ elements.}
                      \item Choose windows with span $s$ centered at each point of estimation, that is, at each $x_i$ or specifically at each redshift, and for each window,
compute the distance to each point of the (local) subset see further below how to choose $s$.
                      \item Find the maximum distance  among the points in the subset and normalize all of them so that the maximum distance becomes 1.
                      \item Assign a weight to each point through the tricube kernel, Eq. (\ref{eq:funw}).
                      \item  Using a linear polynomial, do a weighted least squares fit with each subset of data.
			\item Evaluate the regression functions obtained at the corresponding  $x_i$.
                      \item Connect the resulting (fitted) values with a line. {\it This gives a local polynomial nonparametric regression curve which at the same time provides
 a picture of the general trend.}
\item Go back to step 2(a) until all $j$ values are covered, that is, until all $\lambda$ values are addressed. {\it The output of this process is a set of $\hat{\eta}_{i}(\lambda_{j})$ elements}.
                \end{enumerate}
\item Finish \textit{Simex}:
\begin{enumerate}
\item Select an $i$  value.
\item Perform a standard quadratic polynomial regression with all the
$\hat{\eta}_{i}(\lambda_{j})$ elements in the selected column,
see Eq. (\ref{Eq:polyn}). {\it This will give $\hat{\eta}_{i}$ as a function of arbitrary $\lambda$.}
\item Take the $\lambda \rightarrow -1$ limit.
\item Go back to step 3(a) until all $i$ values are covered.
{\it The  result is a vector $\hat{y}_{i}$ which gives us the global trend of the data  in the light of observational errors.}
%{\it The  result is a vector of $\hat{y}_i$ elements has become the matrix of  $\eta_i(\lambda_j)$ elements.}
\end{enumerate}
\end{enumerate}

Now, even though it is true that in \textit{Loess} the degree of polynomial and the weight function can all affect the trade-off between the bias and variance of the fitted curve, the size of the window span has the most important effect. In this work we have 
faced this issue by using the cross-validation method to determine the optimal value of the span. In practice, 
one should first repeat the procedure at step $4$ for different values of $s$ and choose the one which gives us the lowest value of the cross-validation function $CV(s)$, Eq. (\ref{Eq:CV}).

%%%%%%%%%%%%%%%%%%%%%%%%%%

\section{Observational data, results and discussion \label{dataresults}}
\subsection{Hubble parameter data and distance modulus data}
In order to reconstruct the cosmic expansion, we use (as raw data) two popular data sets. In the spirit of the method proposed we do not assume
any cosmological model, i.e. just let the method smooth out the data without using the fact they are astronomical data which should accommodate some
known physical behaviour. Therefore, in the next subsection we strictly follow the recipe outlined before and perform local polynomial reconstructions of the global trend and apply
the necessary steps to infer the optimal values of the constants of the polynomials. Later in this section we perform a usual MCMC cosmological fit by considering
a specific global form of the dark energy equation of state (EoS) parameter $w(z)=p/\rho$ and obtain the confidence intervals by the usual error propagation technique.
%The results are compared graphically. 
We then identify some significant redshift values and compare the results (see Tables \ref{Table:2} and \ref{Table:3}) obtained from our cosmological-model-independent (non-parametric) technique and from the 
usual parametric approach (based in this case on the CPL scenario).

But let us first provide for completeness some details about the data themselves.
 The first data set we consider is the compilation of Hubble parameter measurements estimated with the differential evolution 
of passively evolving early-type galaxies in the redshift range $0<z<1.75$ recently updated in \citep{Jimenez12} but first reported in \citep{Jimenez02}. 
The main idea supporting this approach is the measurement of the differential age evolution of these chronometers as a function of redshift, which provides 
a direct estimate of the Hubble parameter $H(z) = -1/(1 + z)dz/dt \simeq -1/(1 + z)\Delta z/\Delta t$. The main strength of this approach is the confidence 
on the measurement of a differential quantity, $\Delta z/ \Delta t$, which provides many advantages in minimizing many common issues and systematic effects, 
besides this approach furnishes a direct measurement of the Hubble parameter, and not of its integral, in contrast to SNe Ia or angular/angle-averaged BAO. So, 
we can use the direct measurements of the Hubble parameter to reconstruct the cosmic expansion.

On the other hand, the Hubble diagram, which is a plot of apparent fluxes (usually expressed as magnitudes) of some types of objects at cosmological distances, 
against their redshifts, was initially introduced as a way to demonstrate the expansion of the Universe \citep{Riess:1998cb,Perlmutter:1998np}, and subsequently 
to determine the expansion rate (that is to say, the Hubble constant $H_0$) so to reconstruct the trend of the Hubble diagram from observational data turns out to 
be vital.  Here we construct the Hubble diagram using the updated compilation released by the Supernova Cosmology Project (SCP): the Union2.1 compilation \citep{Union21}. 
The Union2.1 compilation, made up of 580 data points, is the largest published and spectroscopically confirmed SNIa sample to date. 

Because the data points of Union2.1 are given in terms of the distance modulus $\mu_{obs}(z_i)$, we can use them in principle to reconstruct the Hubble diagram in a quite direct way. Nevertheless due to the fact that their covariance matrix is not diagonal one could not  use our method without betraying one of its basic assumptions (decorrelation). Thus, in order to perform the Simex method, we have to decorrelate the data so as to work with new quantities  with diagonal covariance matrix. As suggested in \citep{Whiten}, by solving the eigenvalue problem, the ``decorrelated'' diagonal covariance matrix (which we stress to be different from the ``observational'' diagonal covariance matrix given in the Union2.1 website) can be found. Besides, with the transpose of the matrix that diagonalizes the covariance matrix, the $\mu_{obs}(z_i)$ vector can be also transformed into a new quantity (``decorrelated'' $\mu_{obs}(z_i)$) which becomes the input for our method (i.e., the first step of Simex). Finally, after having added the errors taken from our ``decorrelated'' diagonal covariance matrix to this new ``decorrelated''  $\mu_{obs}(z_i)$, we transform back our results into their original form, i.e. we go back to $\mu_{obs}(z_i)$ and go on with other steps. In what follows, when we refer to the results from Type Ia Supernovae data, remember that we have worked with decorrelated data.

\begin{figure*}
\begin{tabular}{cc}
\includegraphics[width=0.48\textwidth]{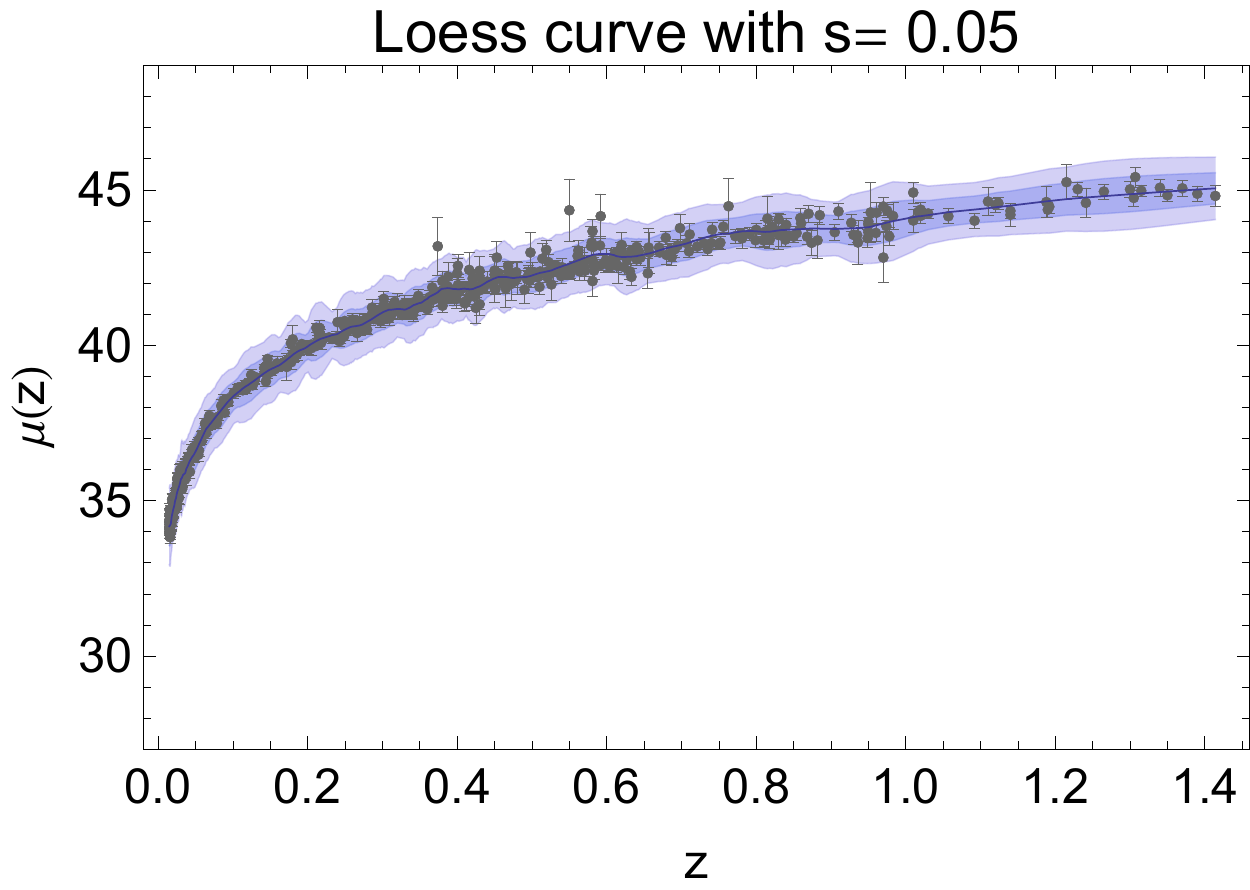}&
\includegraphics[width=0.48\textwidth]{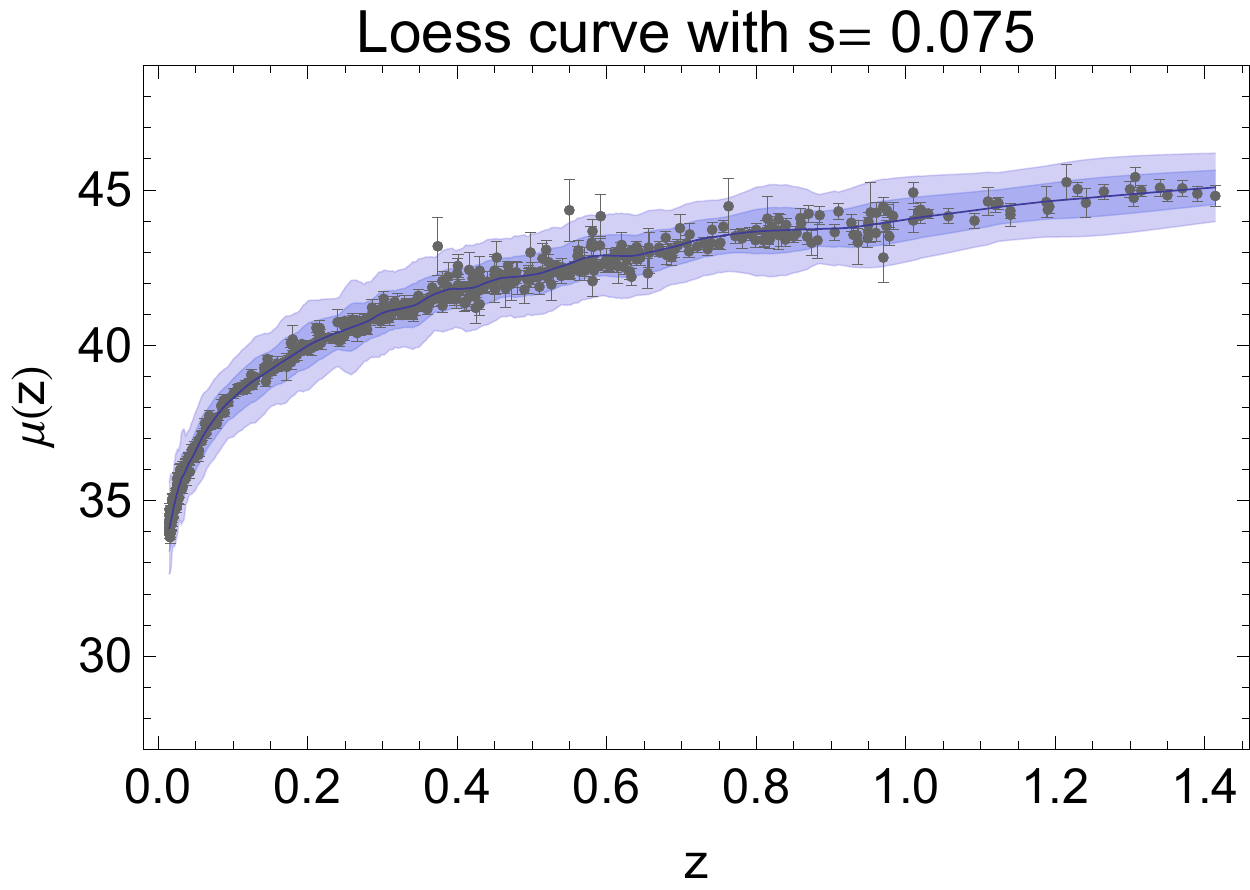}\\
%(a) & (b)
\end{tabular}
\begin{tabular}{cc}
\includegraphics[width=0.48\textwidth]{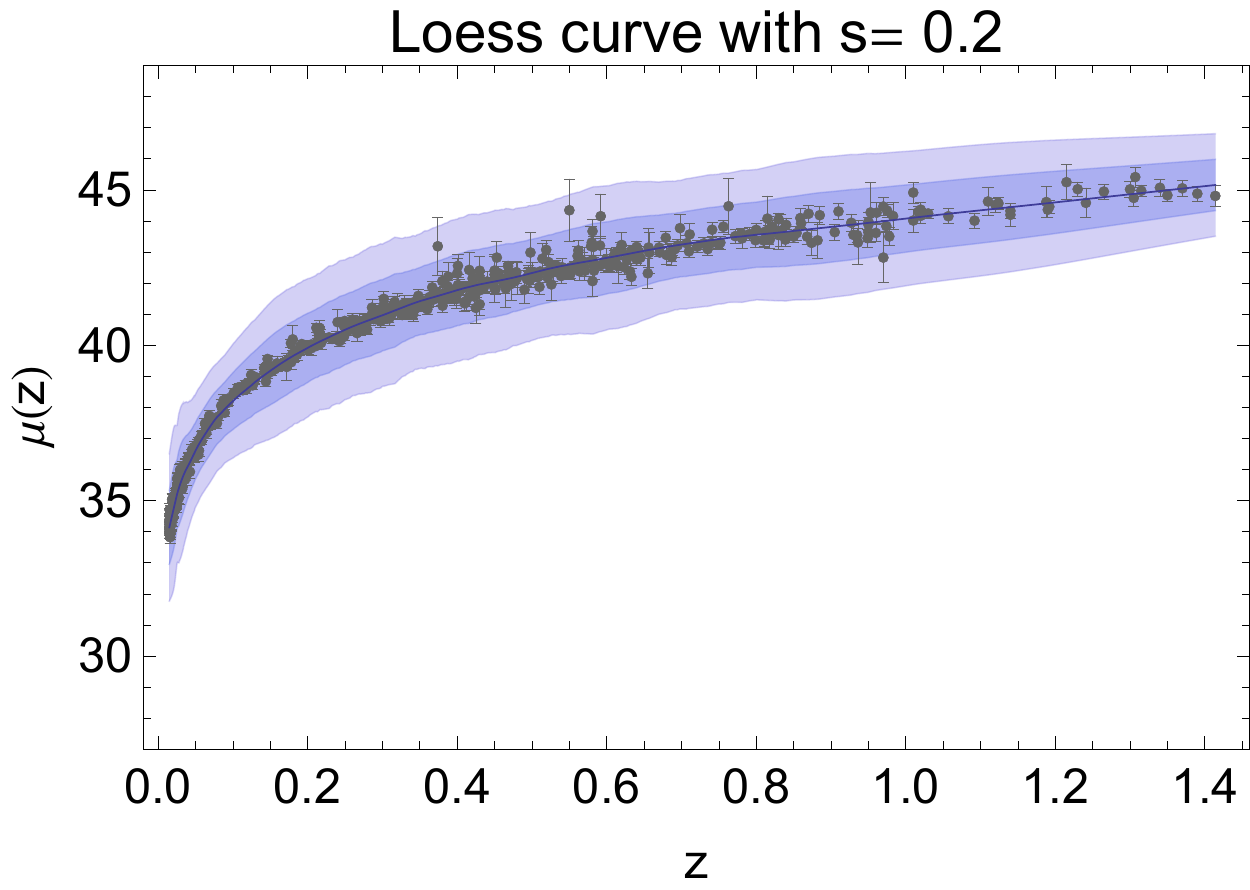}&
\includegraphics[width=0.48\textwidth]{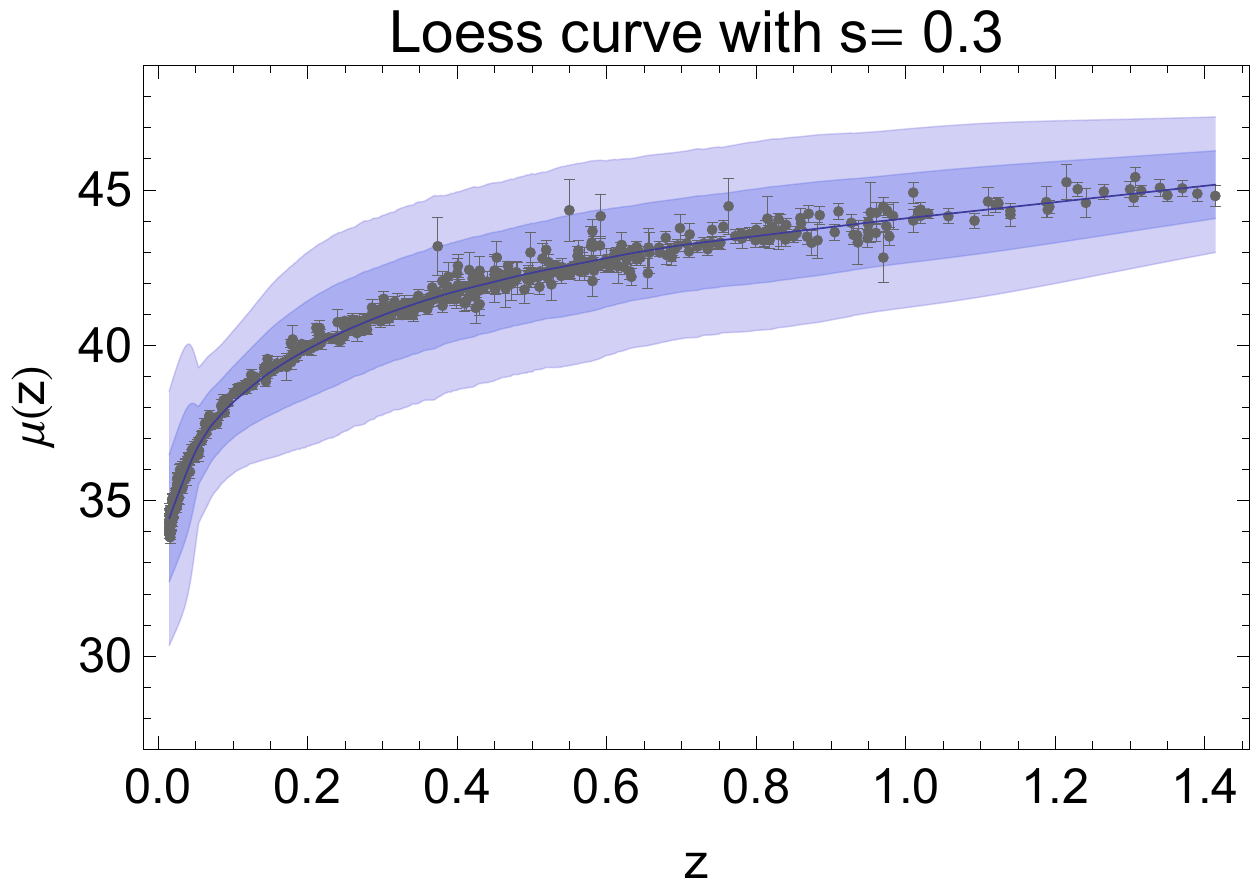}\\
%(c) & (d)
\end{tabular}
\caption{\label{Fig:Loess4}  \textit{Loess} plots with different bandwidths. The gray points are the moduli distances, including uncertainties, of the Type Ia Supernovae. For greater clarity of the general trend, here is not shown the blue points representing the simulated data obtained from our \textit{Loess}+\textit{Simex} method, instead of that, we just show the central line that is obtained just connecting this simulated data.  The shaded contour represent $1\sigma -2 \sigma$ confidence level.}
\end{figure*}

\begin{figure}
\includegraphics[width=0.4\textwidth]{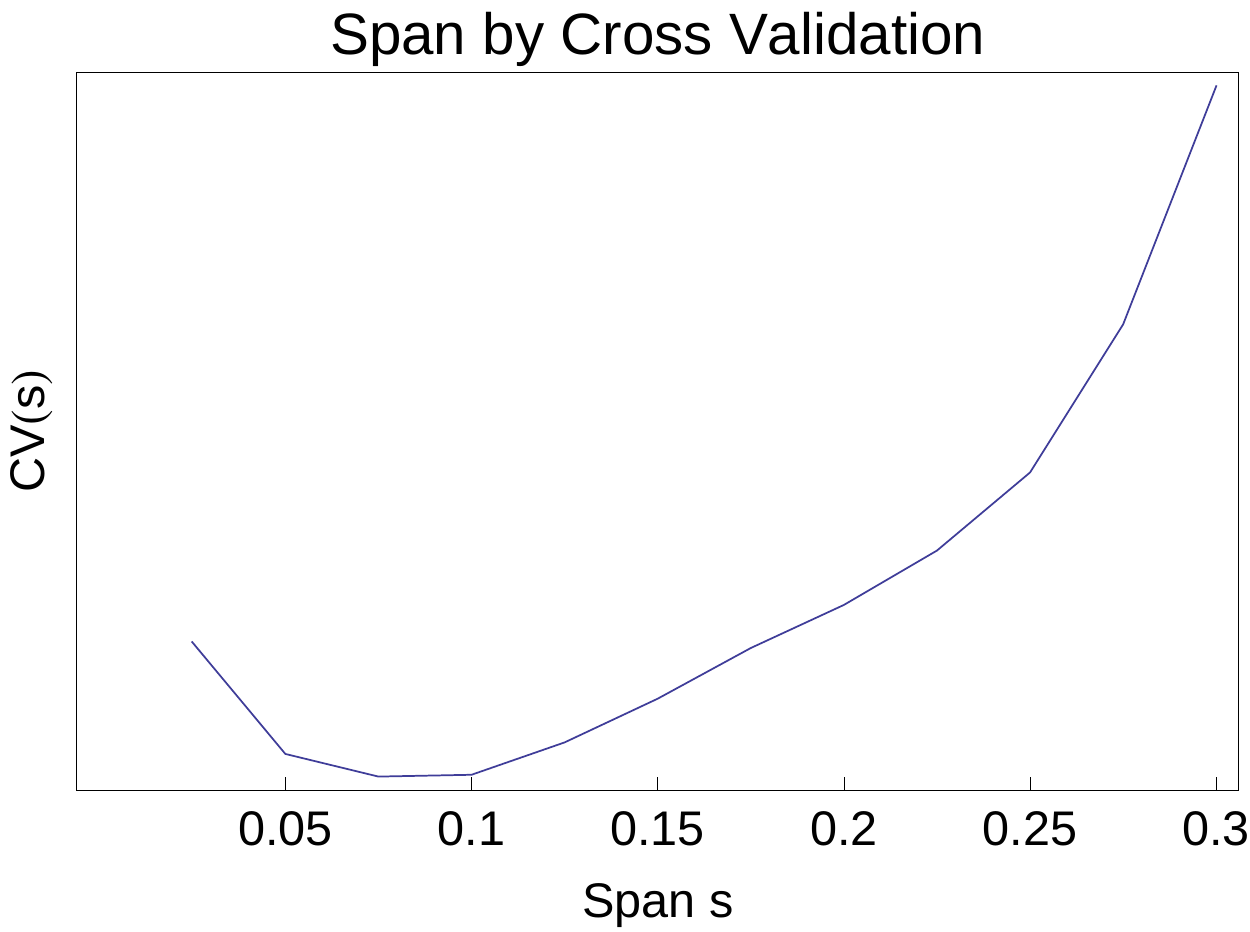}
\caption{\label{Fig:CVSNIa} Plot of $CV(s)$ versus $s$ for Type Ia Supernovae. In this case, a good choices for $s$ could be 0.075 or 1.0. Notice that for this data sample, the range of values of $s$ that miminizes the $CV(s)$ function is clearly determined. }
\end{figure}

\subsection{Results and discussion\label{discussion}}

To obtain the best span for Hubble parameter measurements via cross-validation, we chose subsets containing $30,35,40,...,100 \%$  of the
data corresponding to $s =0.3,0.35,0.4,...,1.0$, respectively. We conclude that $s=0.9$ is the optimal span. 

In Figure \ref{Fig:Loess2} the most illustrative results coming from our \textit{Loess}+\textit{Simex} factory are displayed. Notice we are showing the \textit{Loess}+\textit{Simex} curve with $s=0.2$ just to highlight the effect of a too small span.

In this figure the respective $68\%$ and $95\%$ confidence regions surrounding the reconstructed curves are also shown.

The equivalent degrees of freedom,  $df_{mod}$, which would be the number of parameters of the fit if we were doing a parametric regression, have been computed using $df_{mod}=\mathrm{Tr} (\mathbf{S}\mathbf{S}^T) $. In Table \ref{Table:1} we 
present the values for the equivalent degrees of freedom of the regression, as well as the comparison with the ones obtained with the alternative definition $df_{mod}=\mathrm{Tr} (\mathbf{S}) $. In the case of a too small span ($s=0.2$), the regression curve is a kind of zigzag curve and the 
equivalent degrees of freedom are around 8, which makes sense due the high response to the fluctuations in the data and because the regression curve is susceptible to capture the random error in them. On the other hand, when the span is large ($s=1.0$), the equivalent degrees of freedom are approximately the same as for a linear parametric model, which can be understood because with wider fitting windows the observations tend to cancel each other having less influence on local 
regressions.

{\renewcommand{\tabcolsep}{1.mm}
{\renewcommand{\arraystretch}{1.}
\begin{table}[htbp]
\begin{minipage}{0.5\textwidth}
\caption{\label{Table:1} Equivalent number of parameters or equivalent degrees of freedom $df_{mod}$ for the regression curve obtained from \textit{Loess}+\textit{Simex} factory using Hubble parameter and Supernovae measurements. Notice that despite the values obtained from the two definitions are not equal, they are of similar magnitude.}
\centering
\resizebox*{0.75\textwidth}{!}{
\begin{tabular}{ccc|ccc}
\hline \hline
\multicolumn{3}{c}{Hubble data} &  \multicolumn{3}{|c}{Supernovae}\\
\hline
Span   &  \multicolumn{2}{c|}{$df_{mod}$} & Span   &  \multicolumn{2}{c}{$df_{mod}$} \\
& $\mathrm{Tr} (\mathbf{S})$ & $\mathrm{Tr}(\mathbf{S} \mathbf{S}^T)$ &  & $\mathrm{Tr}(\mathbf{S})$ & $\mathrm{Tr}(\mathbf{S}\mathbf{S}^T)$ \\
\hline
$0.2$      & $9.46$    & $8.22$   & $0.05$     & $35.40$   & $29.20$   \\
$0.4$      & $4.09$    & $3.44$   & $0.075$      & $23.34$   & $19.24$    \\
$0.9$      & $1.79$    & $1.51$   & $0.2$      & $9.03$    & $7.33$   \\
$1.0$      & $1.62$    & $1.37$   & $0.3$      & $5.85$    & $4.80$   \\
\hline \hline
\end{tabular}}
\end{minipage}
\end{table}}}

{\renewcommand{\tabcolsep}{1.mm}
{\renewcommand{\arraystretch}{1.5}
\begin{table*}
\caption{$1\sigma$ confidence levels for Hubble data. \textit{Loess}+\textit{Simex}: only diagonal statistical errors and $s=0.9$. Data fit: full covariance matrix. \textit{Planck}: joining \textit{Planck} and \textit{WMAP9} CMB data (for CPL the team does not give errors on parameters, but only $95\%$confidence limit values). \textit{WMAP9:} joining \textit{WMAP9} with SPT+ACT CMB data, SNLS SN, lensing and BAO, from parameters table in the official web-site.}\label{Table:2}
\begin{tabular}{c|ccccc}
\hline\hline
$z$ & \textit{Loess}+\textit{Simex} & data fit & \textit{Planck} & \multicolumn{2}{c}{\textit{WMAP9}}\\
    &       & CPL      & $\Lambda$CDM    & $\Lambda$CDM & $w= $const.        \\
\hline\hline
$0.18$  &$(65.05;79.20)$ & $(53.36;105.33)$ & $(72.14;75.44)$ & $(73.90;81.37)$ & $(71.99;78.96) $ \\
                       % & & $ $ & $ $ & $ $ & $ $ \\
\hline
$0.25$&$(68.52;81.98)$ & $(51.73;112.68)$ & $(74.87;78.57)$ & $(75.84;84.32)$ & $(73.59;82.28)$ \\
                       % & & $(0.773;1.292)$ & $(1.133;1.147)$ & $(1.080;1.127)$ & $(1.056;1.138)$ \\

\hline
$0.3$  &$(71.26;83.94)$ & $(52.15;118.04)$ & $(76.96;80.94)$ & $(77.38;86.57)$ & $(74.96;84.85) $ \\
                       % & & $ $ & $ $ & $ $ & $ $ \\
\hline
$0.35$  &$(73.99;85.89)$ & $(53.55;123.48)$ & $(79.15;83.43)$ & $(79.06;88.93)$ & $(76.51;87.56)$ \\
                      %  & & $ $ & $ $ & $ $ & $ $ \\

\hline
$0.4$  &$(76.82;88.68)$ & $(55.62;129.00) $ & $(81.44;86.03) $ & $(80.86;91.40) $ & $(78.24;90.41) $ \\
                       %% & & $ $ & $ $ & $ $ & $ $ \\

\hline
$0.5$ &$(82.47;94.34)$ & $(60.86;140.28)$ & $(86.32;91.55)$ & $(84.83;96.66)$ & $(82.18;96.47)$ \\
                       % & & $(0.910;1.609)$ & $(1.306;1.336)$ & $(1.208;1.292)$ & $(1.179;1.334)$ \\
\hline
$0.75$ &$(94.47;108.19)$ & $(76.36;169.83)$ & $(100.13;107.07)$ & $(96.67;111.52)$ & $(94.50;113.43)$ \\
                       % & & $(1.141;1.948)$ & $(1.515;1.563)$ & $(1.377;1.491)$ & $(1.356;1.569)$ \\

\hline\hline
\end{tabular}
\end{table*}}}

\begin{figure}
\includegraphics[width=0.48\textwidth]{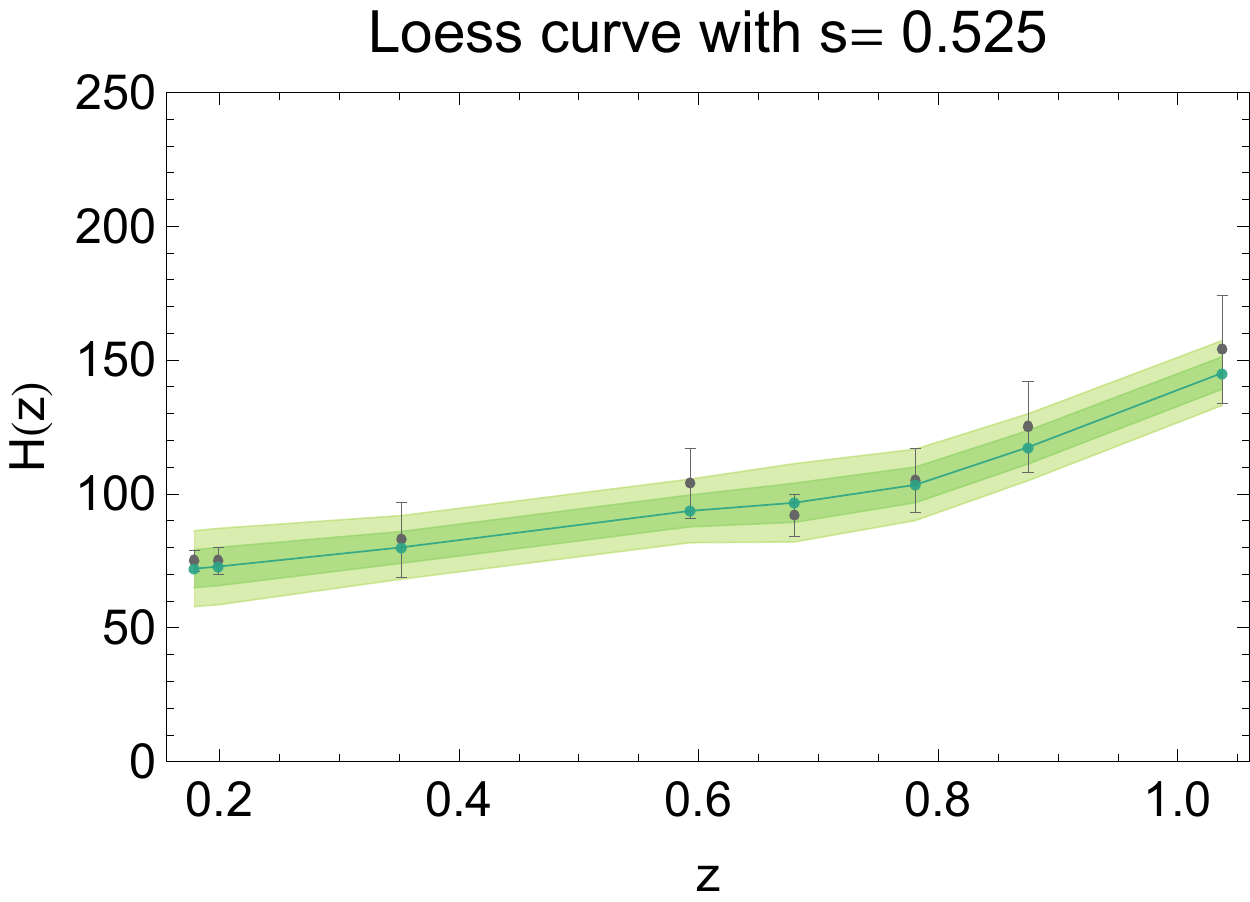}
\caption{\label{Fig:JimenezR6} \textit{Loess} plots with different bandwidths. The gray points are the measurements of Hubble parameter including their uncertainties from \citep{Moresco:2012jh}, 
the green points are the simulated data resulting from our \textit{Loess}+\textit{Simex} method. The central green line is obtained just connecting this dots and 
represents the best fit. The shaded contour represent $1\sigma -2 \sigma$ confidence level for our best fit.}
\end{figure}

\begin{figure}
\includegraphics[width=0.48\textwidth]{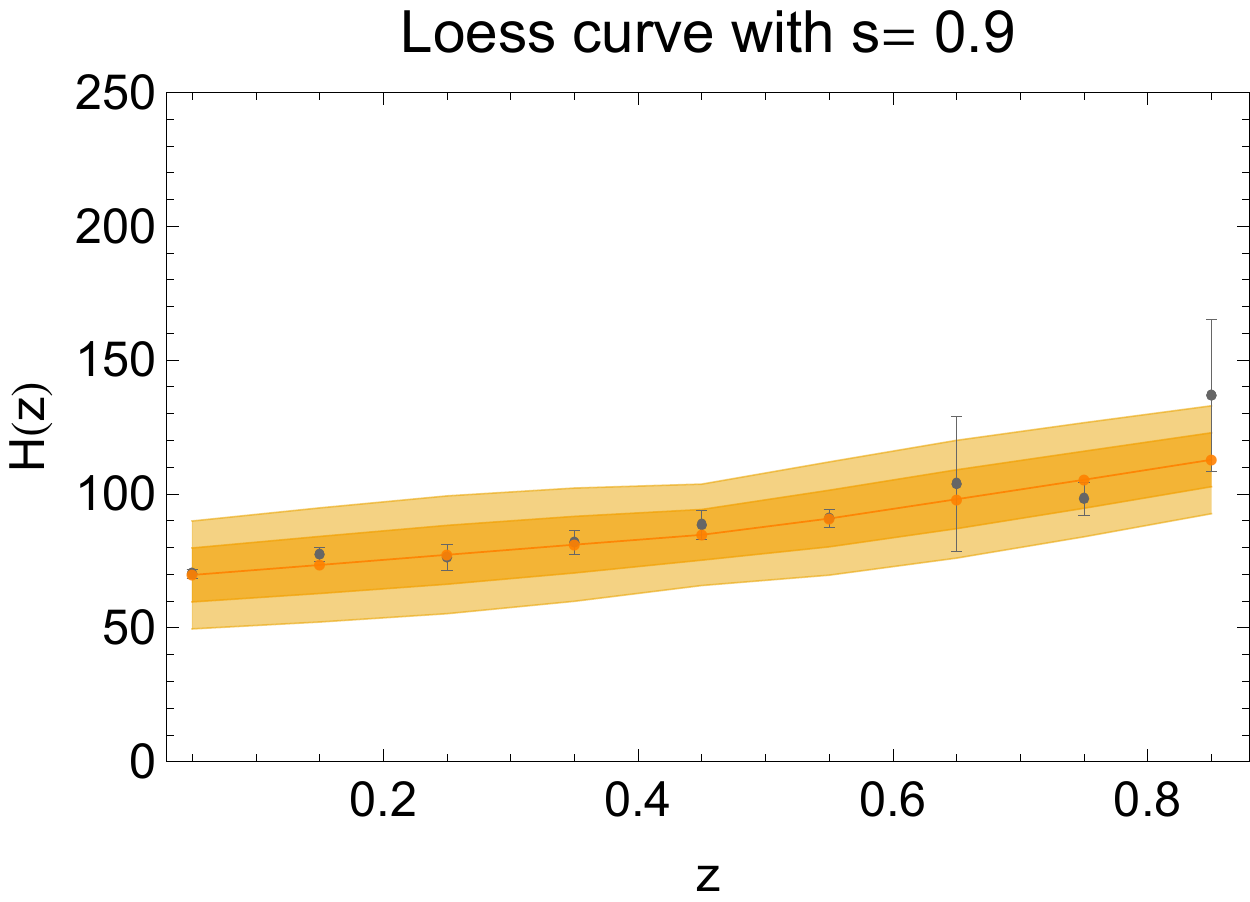}
\caption{\label{Fig:WiggleZ} \textit{Loess} plots with different bandwidths. The gray points are the measurements of Hubble parameter including their uncertainties from \citep{Blake:2012pj}, 
the orange points are the simulated data resulting from our \textit{Loess}+\textit{Simex} method. The central orange line is obtained just connecting this dots and 
represents the best fit. The shaded contour represent $1\sigma -2 \sigma$ confidence level for our best fit.}
\end{figure}

Moreover, as can be seen from Figure \ref{Fig:Loess2}, the choice of the window width (span) has an important effect. In this case, a span that is too small (means that insufficient data fall within the window) results in a large variance, as we can see for the cases with $s=0.2$ or $s=0.4$, although the latter in a lesser degree. On the other hand, if the span is too large, the data will be 
over-smoothed, resulting in a loss of important information and, consequently, bias in the fitted curve and large confidence regions, see the figure with $s=1.0$ (the cent percent of the data).

Figure \ref{Fig:CVHz} shows the $CV(s)$ function versus $s$ for Hubble parameter measurements. $s=0.9$ turns out to be the value of $s$ that minimizes the cross-validation function providing  a compromise between the over-fitting of the last panel in Figure \ref{Fig:Loess2}, and the lack of fit that occurs in the upper two panels of the same figure, besides of a slightly tight confidence region.  Although Figure \ref{Fig:CVHz}  provides little clarity to select directly from it an appropiate value of the span, it suggests that $s$ should be larger than $s=0.8$.

In Figure \ref{Fig:Loess4}, we present the reconstruction of the Hubble diagram (blue line), using the supernovae data, as well as the respective confidence regions. The original data, as well as their measurement errors are drawn in gray. As in the case of Hubble parameter measurements, the results displayed come from the \textit{Loess}+\textit{Simex} factory. The main difference between Type Ia Supernovae data and Hubble parameter measurements is the amount of data points available. It is natural to expect that a large amount of 
data will need a different value of the span as compared to the one chosen for a sample such as the Hubble parameter measurements.
For supernovae we have explored $s= 0.025,0.05,0.075,...,0.3$, that is to say, we have used the $2.5,5,7.5,...,30\%$ of the data to obtain the best value of the span via cross-validation, and we have kept the same values for $\lambda$ as in the case of Hubble parameter measurements. As we will discuss later, the best value of $s$ is $s=0.075$.  In  Figure \ref{Fig:Loess4} we present some representative results for which several span choices have been considered.

From Figure \ref{Fig:Loess4} we can immediately see that the trend of the reconstructed Hubble diagram approaches the curve that would result from a standard least 
squares fit if a large window width is used, see \textit{Loess} curves with $s=0.3$ and $s=0.2$. However, what we would like is to have a curve as smooth as 
possible that reproduces faithfully the behavior of the data but without oversmoothing. Thus, to choose the best value of the span turns out to be vital.
From cross-validation (see Figure \ref{Fig:CVSNIa}) we have found that as already mentioned the best choice is $s=0.075$. In the plot of $CV(s)$ versus $s$, the region that corresponds to the best value of  $s$ is quite broad and rather flat, thus from Figure \ref{Fig:CVSNIa} we can easily identify that the optimal values of $s$ are between $s=0.075$ and $s=0.1$. The value of $s$ that minimizes the cross-validation function is $s=0.075$. Additionally, in Table \ref{Table:1} we present the values of the equivalent degrees of freedom obtained from $df_{mod}=\mathrm{Tr}
 (\mathbf{S}) $ and $df_{mod}=\mathrm{Tr} (\mathbf{S} \mathbf{S}^T) $. In this case the sample is larger than the one of Hubble parameter measurements, thus the equivalent number of parameters of the nonparametric regression is much larger. The same reasoning as in the case of Hubble parameter measurements is followed for Type Ia Supernovae data regarding the equivalent degrees of freedom of the nonparametric  regression: the $df_{mod}$ is larger when $s$ is small and, $df_{mod}$ is smaller when $s$ is large.
 
Finally, to test the reliability and robustness of the \textit{Loess}+\textit{Simex} method, in Tables \ref{Table:2} and \ref{Table:3} we present a comparison between predictions for the measurements of the Hubble parameter and the moduli distance from \textit{Loess}+\textit{Simex} and from conventional MCMC cosmological fits by considering a specific global form of the EoS parameter $w(z)$. 

The compilation of Hubble parameter measurements reported in \citep{Jimenez12}, which we have used throughout our analysis, is composed by three sub-samples: the first one reported in \citep{Simon:2004tf}, the second one reported in \citep{Stern:2009ep} and the third one reported in \citep{Moresco:2012jh}.  Our guess is that, probably, the latest compilation \citep{Moresco:2012jh} is the most reliable. So, we perform the reconstruction of cosmic expansion with this sub-sample. In Table \ref{Table:2}  we present the results coming from our \textit{Loess}+\textit{Simex} factory by using the Hubble parameter measurements reported in \citep{Moresco:2012jh}.  In this case, as can be seen in Table \ref{Table:2}, the results from our proposal are much better in comparison with the ones obtained from a standard MCMC cosmological fit for the CPL scenario and almost as good as the results for the $\Lambda$CDM model by adopting the Planck and WMAP9 CMB data. For this sub-sample the best span is $s=0.525$ and the equivalent degrees of freedom are $ df_{mod}=2.47$; the respective reconstructed curve can be seen in Figure \ref{Fig:JimenezR6}. 

The previous result lead us to think that the \textit{Loess}+\textit{Simex} method is quite sensitive to the quality of the data. To confirm our suspicion, we use the Hubble parameter measurements recently reported by the WiggleZ Dark Energy Survey \citep{Blake:2012pj}. For this sample, the optimal span and the equivalent degrees of freedom turn out to be $s=0.9$ and $ df_{mod}=1.41$, respectively.  The nonparametric reconstruction of the cosmic expansion by using this compilation, which can be seen in the Figure \ref{Fig:WiggleZ}, together with the results coming from Type Ia Supernovae data (see Table \ref{Table:3}), allow us to conclude that the \textit{Loess}+\textit{Simex} factory is a very promising approach to reconstruct global trends in a non-parametric way if one have observational data of high quality.
 
 {\renewcommand{\tabcolsep}{1.mm}
{\renewcommand{\arraystretch}{1.5}
\begin{table*}
\caption{$1\sigma$ confidence levels for Union2.1 SN data. \textit{Loess}+\textit{Simex}: diagonal covariance matrix and $s=0.075$. Data fit: full covariance matrix. \textit{Planck}: joining \textit{Planck} and \textit{WMAP9} CMB data (for CPL the team does not give errors on parameters, but only $95\%$confidence limit values). \textit{WMAP9:} joining \textit{WMAP9} with SPT+ACT CMB data, SNLS SN, lensing and BAO, from parameters table in the official web-site.}\label{Table:3}
\begin{tabular}{c|ccccc}
\hline\hline
$z$ & \textit{Loess}+\textit{Simex} & data fit & \textit{Planck} & \multicolumn{2}{c}{\textit{WMAP9}}\\
    &       & CPL      & $\Lambda$CDM    & $\Lambda$CDM & $w= $const.        \\
\hline\hline
$0.05$  &$(35.975;37.221)$ & $(36.706;36.765)$ & $(36.740;36.742)$ & $(36.743;36.754)$ & $(36.745;36.762)$ \\
\hline
$0.1$  &$(37.671;38.911)$ & $(38.254;38.383)$ & $(38.318;38.323)$ & $(38.325;38.347)$ & $(38.327;38.361)$ \\
\hline
$0.25$ &$(39.825;41.170)$ & $(40.360;40.704)$ & $(40.497;40.511)$ & $(40.515;40.565)$ & $(40.511;40.594)$ \\
\hline
$0.5$  &$(41.612;42.924)$ & $(42.025;42.575)$ & $(42.246;42.271)$ & $(42.281;42.364)$ & $(42.258;42.406)$ \\
\hline
$0.75$ &$(42.928;44.087)$ & $(43.033;43.645)$ & $(43.306;43.341)$ & $(43.357;43.462)$ & $(43.314;43.505)$ \\
\hline
$1$    &$(43.361;44.726)$ & $(43.761;44.389)$ & $(44.069;44.112)$ & $(44.132;44.253)$ & $(44.073;44.291)$ \\
\hline\hline
\end{tabular}
\end{table*}}}

%%%%%%%%%%%%%%%%%%%%%%%%%%%%%%%%%%%%%%%%%%%%%%%%%%%%%%%%%%%%%%%%%%%%%%%%%%%%%%%%%%%%%%%%%%%%%%%%%%

\section{Conclusions\label{conclusions}}

The goal of this work was to reconstruct the cosmic expansion of the Universe in a cosmological model independent way through the implementation of a combination of the \textit{Loess} and \textit{Simex} methods. The first one allowed us to obtain smoothed curves of the general trend via a nonparametric regression and the second 
one addresses the fact that the effect of measurement error on a variable can be determined via simulation. In general, we can say that our proposal, the \textit{Loess}+\textit{Simex} factory, grasps successfully the global trend underlying the data, the current $H(z)$ measures and the distance moduli of Type Ia Supernovae data, taking into account not only the observational measurements but also their error, thus providing a faithful reconstruction of cosmic expansion.

One of the most appealing features of our method is that it can be a valuable technique for visualizing complex relationships and validating models if needed. In Cosmology this feature turns out to be very convenient: since we are interested in gaining as much information as possible from observational data to dilucidate the nature of DE, the \textit{Loess}+\textit{Simex} factory, which is a very simple method that ignores any assumption about the relationships between variables and cosmological quantities,  could become a very promising tool to establish trends and clarify the functional relationship between them. Furthermore, note that our proposal overcomes the issues we list at the beginning of this work:
\begin{itemize}
 \item it does not assume any prior or a initial guess cosmological model and so one avoids possible biases or not quite right assumptions;
 \item it allows to have a first glance at the global trend of data without having to resort to heavy calculations;
 \item it presents the same efficiency along all redshift ranges because it lets the window width vary throughout the redshift range so that one has
 the same number of data points in each fitting window,  this number having been chosen after an optimality test;
 \item the estimation and propagation of the error can be done in a quite direct way and the computational cost is very low.
\end{itemize}
On the other side, the exploration of this approach has also suggested using the method as a very accurate technique for predicting new data points through the local polynomial. This local polynomial is obtained for each subset of data, which is used to compute the final value of the regression function in the corresponding estimation point, but it would also allow
to generate new synthetic points of data where the original sample is poor. Indeed, this feature is linked to the value of the span $s$, such that to obtain mock data, certainly choosing small values of the span $s$ is the best option, because in this way local trends are captured.

Even though our results are very reasonable, it can be clearly noticed that the method is dependent on the choice of the span (or equivalently, on the window width). In the case of small samples, as $H(z)$ data, the use of larger fitting windows leads to more robust fits; in contrast, using smaller fitting windows allows  finding local properties of the trend with a larger redshift resolution. In the case of Type Ia Supernovae, i.e. large samples,  the contrary happens and a lower value of the span is necessary to achieve good results. Besides, and as expected, it can be seen that the election of the span, also introduces a bias in the results when an appropriate value of it has not been chosen. This issue can be faced successfully by implementing the cross-validation method to select the optimal span and, even though it may happen that the method provides little help for selecting the span, it can in principle suggest an interval of the best values of $s$ to produce a smooth curve. Thus, \textit{Loess}+\textit{Simex} along with the cross-validation method turns out to be a self-sufficient approach in the sense we do not have to choose anything by hand to obtain smooth global trends.

On the other hand, from the results presented in Tables \ref{Table:2} and \ref{Table:3}, we have noticed that our approach is sensitive to the quality of the data. Thus, if the data sample contains data points of poor quality, the \textit{Loess}+\textit{Simex} factory will produce smooth curves but with broad confidence regions. Thus, one could expect that in the opposite case, that is to say, with more data of high quality, such as the Type Ia Supernovae data, our proposal could be a quite reliable tool for reconstructing global trends of cosmological data.

Our results are meant primarily to illustrate the method and to suggest that the \textit{Loess}+\textit{Simex} factory has good prospects to reconstruct the expansion history of the Universe. The strength of the approach relies in that it is a model-independent  and nonparametric method, as does not assume any prior nor the energy contents of the Universe or some other property related to a cosmological model. Besides, our method allows to draw the confidence regions around the regression curve, and although they seem to be broader than those one could obtain by using other methods, we have made sure that we have not underestimated them. Thus, we believe it offers an alternative 
way to study cosmological data in order to find possible parametrizations that reliably describe the data with no prior knowledge of a cosmological model.

Finally, since we are convinced that one can gain useful information from approaches that can reconstruct the properties of DE or the history of the expansion rate, our next step, which we leave for future work,  would be to test the power of our method to reconstruct derived quantities, as well as the ability of our method to discriminate between dark energy models that in principle represent the underlying structure of the data.

%**********************************************************************

\begin{acknowledgments}
AM acknowledges financial support from Conacyt-M\'exico, through a PhD grant. RL, VS and IS are supported by the Spanish Ministry of Economy and Competitiveness through research projects FIS2010-15492 and Consolider EPI CSD2010-00064, and also by the Basque Government through research project   GIC12/66, and by the University of the Basque Country UPV/EHU under program UFI 11/55. C. Escamilla-Rivera is supported by Fundaci\'on Pablo Garc\'ia-FUNDEC, M\'exico.
\end{acknowledgments}

\bibliography{biblio}
\bibliographystyle{apsrev} %unsrtnat
%is-abbrv

\end{document}